\documentclass[ a4paper,10pt, fleqn]{scrartcl}

\usepackage[utf8]{inputenc}
\usepackage{ngerman}
\usepackage{url}
\usepackage{graphicx}
\usepackage{natbib}
\usepackage{scrpage2}
\usepackage{hyperref} 
\usepackage{amsmath}
\usepackage{amsfonts}
\usepackage{amssymb}
\usepackage{amsthm}
\usepackage{setspace}
\usepackage[normal,small]{caption}
\usepackage[switch,pagewise]{lineno}

\modulolinenumbers[5]

\cohead{\normalfont \small \scrauthor: \scrtitle}

\newcommand{\vect}[1]{\boldsymbol{#1}}

\def\aj{AJ}%
\def\apj{ApJ}%
\def\apjs{ApJS}%
\def\aap{A\&A}%
\def\mnras{MNRAS}%
\def\pasj{PASJ}%
\areaset[current]{18.2cm}{25cm}

\begin{document}
\newcommand{\scrauthor}{Rapha\"el Errani}
\newcommand{\scrtitle}{Modell zur Entstehung eines Rings aus Dunkler Materie in der Milchstraßenebene} 
\newcommand{\scrdate}{\today}

\begin{titlepage}
%

\fbox{
\begin{tabular}[b]{llr} 
	\includegraphics[height=1.2cm]{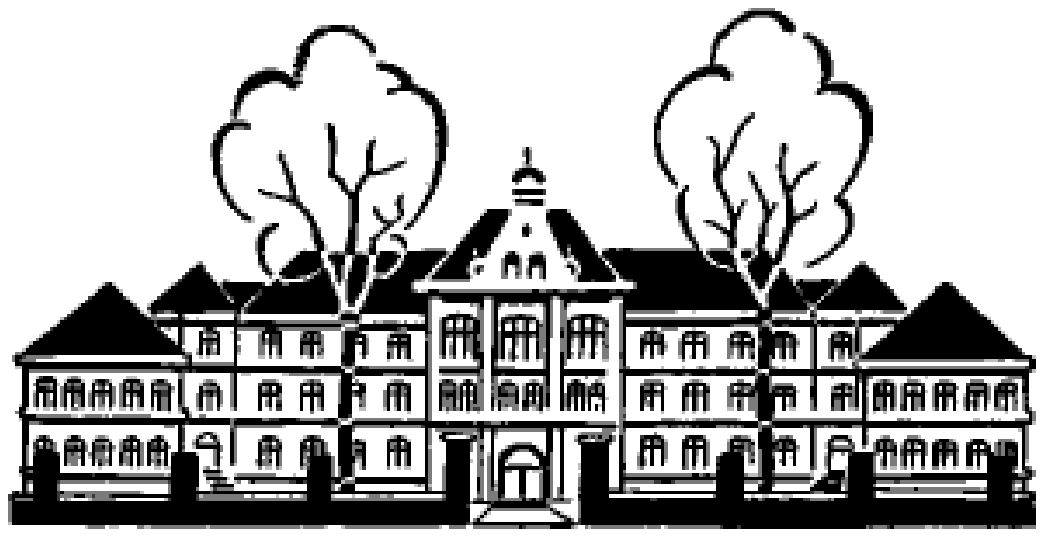}&
	\begin{tabular}[b]{l}
	\textsf {Athenaeum Stade} \\
	\textsf {Harsefelder Straße 40} \\
	\textsf {21680 Stade}
	\end{tabular}
  \hspace{7.0cm} &
\begin{tabular}[b]{l}
	\textsf {Druckdatum}\\
  	\textsf {\scrdate}
\end{tabular}
\end{tabular}
}

\begin{center}
 \begin{minipage}[t]{16.5cm}

	\begin{center}
		\vspace{2.5cm}
		 \textbf{\textsf {\huge \scrtitle}}\\
		\vspace{1cm}
		{\large \scrauthor \small{$^{~1}$}\\}

\vspace{0.85cm}

 \begin{minipage}[t]{15.0cm}

{\small{
\begin{tabular}[t]{ll}
$^{1}$\hspace{-10pt} &AG für Astronomie am Athenaeum Stade\\
	  &e-mail: \href{mailto:Raphael.Errani@gmx.de}{Raphael.Errani@gmx.de}
\end{tabular}

\vspace{0.6cm}
Angenommen zur Veröffentlichung in der\\
\textit{Jungen Wissenschaft} am 14. Januar 2010

}}

\vspace{1.cm}
\begin{center} \textsf{ \textbf{Abstract}}\end{center}
\vspace{0.1cm}


Die Verteilung von Wasserstoffgas sowie die Rotationskurve der Milchstraße deuten auf die
Existenz eines Rings aus kalter Dunkler Materie innerhalb der Milchstraßenebene in einem Abstand von rund 14 Kiloparsec um das galaktische Zentrum.
Es wird mit Hilfe eines semianalytischen Modells die Entstehung eines solchen Rings durch den
Einfall von Zwerggalaxien in die Milchstraße simuliert.
Die Entstehung des Rings kann sowohl durch einen Einzeleinfall als auch durch den Einfall mehrerer Zwerggalaxien erklärt werden, solange die numerische Exzentrizität der Orbits nicht 0.2 übersteigt und die Inklination der Orbits klein ist. Auch ein die Mindestmasse einer an der Ringentstehung beteiligten Zwerggalaxie beschreibender Zusammenhang ist ermittelt worden.

\vspace{0.7cm}
\textbf{\textsf{Stichwörter:}~} Dunkle Materie  -- Milchstraße --  Zwerggalaxien  -- Semianalytisches Computermodell

\end{minipage}

\end{center}
\end{minipage}
\end{center}
\end{titlepage}

\setcounter{page}{1}
\tableofcontents
\twocolumn
\section{Einleitung}
\label{sec:einleitung}
Die unter Betrachtung der direkt beobachtbaren Materie zu hohen Geschwindigkeiten von Sternen in Galaxien und von Galaxien in Galaxienhaufen führten zur Annahme, dass sich in Galaxien und Galaxienhaufen mehr als nur die direkt beobachtbare Materie befinden muss \citep{zwicky73, bosma81, rubin85}. Diese Materie wird als Dunkle Materie\footnote{Ein Überblick über die Erforschung der Dunklen Materie bis 1999 ist von S. van den Bergh gegeben, zu finden bei arXiv:astro-ph/9904251v1} bezeichnet, die Komponente mit Geschwindigkeiten $v \ll c$ als kalte Dunkle Materie (CDM). 
Der Umstand, dass CDM nicht direkt beobachtet werden kann, wird dadurch erklärt, dass sie nicht elektromagnetisch wechselwirkt und dementsprechend keine elektromagnetischen Wellen aussendet. Diese Eigenschaft prägte den Namen der nur schwach wechselwirkenden, aber schweren Wimps (Weakly interacting massive particles). Das Standardmodell kennt keine Teilchen, die diesen Anforderungen entsprechen, es existiert jedoch eine Vielzahl an Theorien, welche die Existenz solcher Teilchen postulieren. Die wohl populärste dieser Theorien ist die der Supersymmetrie.

Aus der Verteilung von Wasserstoffgas in der Milchstraße wurde auf die Massenverteilung in der Milchstraße geschlossen. In einem Abstand von 14kpc zum galaktischen Zentrum liegt diesen Daten nach eine ringförmige Ansammlung kalter Dunkler Materie mit einer Masse von bis zu $2.8\cdot10^{10}$~M$_\odot$ vor \citep{kalberla07}.
Auch die Rotationskurve der Milchstraße deutet auf eine Massenansammlung  bei $R=14$kpc, siehe zum Beispiel \citet{honma97}. Zusätzlich wird bei $R=14$kpc eine Ansammlung von alten Sternen mit sich von anderen Milchstraßensternen unterscheidender Geschwindigkeitsverteilung beobachtet \citep{ibata03}.
Eine häufig diskutierte mögliche Ursache für die Entstehung dieses CDM-Rings bei $R=14$kpc der Masse $2.8\cdot10^{10}$~M$_\odot$ ist, entsprechend dem Modell der hierarchischen Galaxienentstehung, der Einfall einer einzelnen, massiven Zwerggalaxie \citep{crane03}.

Das Ziel dieser Arbeit ist die Entwicklung eines semianalytischen Computermodells, welches die Simulation der Entstehung des äußeren CDM-Rings durch den Einfall von Zwerggalaxien ermöglicht und nutzt, um Aussagen über physikalische Eigenschaften der Zwerggalaxien und deren Orbits machen zu können. 

\subsection{Modellbeschreibung}
\label{SubSec:Modell}
Semianalytische Modelle haben n-Teilchen-Simulationen gegenüber den Vorteil, dass das Verständnis der funktionalen Zusammenhänge des simulierten Systems leichter fällt. Um realistische und aussagekräftige Ergebnisse zu erhalten, müssen alle wesentlichen Prozesse des Modells einzeln beschrieben und in ihrer Wirkung verknüpft werden. Im Folgenden wird das dieser Arbeit zu Grunde liegende Modell erläutert.

Eine Zwerggalaxie befindet sich bei einem Radius $r_i$ zur deutlich schwereren Milchstraße.
Die Zwerggalaxie bewegt sich durch ein Gebiet der Dichte $\rho$ im Gravitationspotential der Milchstraße mit der Anfangsgeschwindigkeit $\vect{v_i}$.  Die Modellierung der Dichteverteilung der Milchstraße ist in Abschnitt \ref{subsec:dichte_milchstr}, die des Gravitationspotentials in Abschnitt \ref{subsec:Gravitationspotential} beschrieben.
Bei der Bewegung im Gravitationspotential der Milchstraße verliert die Zwerggalaxie durch dynamische Reibung Drehimpuls und Bewegungsenergie. Diese Effekte der dynamischen Reibung im Halo werden in Abschnitt \ref{Subsec:DynFriction} behandelt.
Bei jedem Durchgang der Zwerggalaxie durch die galaktische Scheibe kommt es zu Kollisionen zwischen den H1-Atomen des Wasserstoffgases der Zwerggalaxie und dem Wasserstoffgas innerhalb der galaktischen Scheibe der Milchstraße. Dadurch wird das Wasserstoffgas der Zwerggalaxie abgebremst und sammelt sich in der galaktischen Ebene der Milchstraße. Die Kollisionen werden in Abschnitt \ref{subsec:disk} beschrieben. 
Durch die dynamische Reibung spiralisiert die Zwerggalaxie in Richtung des galaktischen Zentrums der Milchstraße. Dabei steigen mit sinkendem Abstand zum galaktischen Zentrum die auf die Zwerggalaxie wirkenden Gezeitenkräfte, die bei einem Abstand $r_t$ zum galaktischen Zentrum zum Zerreißen der Zwerggalaxie führen. Auf die Gezeitenkräfte wird in Abschnitt \ref{subsec:TidalDisruption} näher eingegangen.
Die zerrissene Zwerggalaxie bildet einen Sternstrom (sowohl aus Sternen als auch aus CDM  und Wasserstoffgas bestehend), der sich zu einer ringförmigen Struktur um das galaktische Zentrum entwickelt. Dieser Sternstrom ist aufgrund der ringförmig und weitläufig verteilten Masse unanfällig für dynamische Reibung des Halos, ändert daher seinen Abstand zum galaktischen Zentrum nur noch wenig. 
Bei dem Einfall einer Zwerggalaxie mit geringer Inklination kann sich also im Laufe der Zeit eine innerhalb der galaktischen Ebene liegende, ringförmige Struktur aus Sternen und CDM bilden.

\section{Simulation}
\label{sec:sim}
In diesem Abschnitt wird die Modellierung der einzelnen in Abschnitt \ref{SubSec:Modell} erwähnten Prozesse beschrieben. Des weiteren wird das verwendete Verfahren zur Zeitintegration der Flugbahn in Abschnitt \ref{subsec:numint} vorgestellt.
Die Implementierung der Simulation erfolgt objektorientiert in der Programmiersprache c++, auf diese wird allerdings in dieser Arbeit nicht näher eingegangen.

\subsection{Dichteverteilung in der Milchstraße}
\label{subsec:dichte_milchstr}
Das verwendete Modell der Milchstraße lässt sich in drei Komponenten aufteilen: Die galaktische Scheibe und den zentralen Bulge, welche überwiegend aus Wasserstoffgas bestehen, sowie das Halo aus Dunkler Materie. 

Die Dichteverteilung $\varrho_h(r)$ des sphärisch symmetrischen Halos wird mit Hilfe des von 
\citet{saha09} behandelten Dichteprofils beschrieben, welches für $r \gg r_{c,h}$ dem NFW-Profil \citep{navarro96} entspricht:
\begin{linenomath}
\begin{equation}
 \label{equ:pseudoisothermal}
\rho_h(r) = \rho_{0,h} \cdot \Bigl[1+ \frac{r^2}{r_{c,h}^2}\Bigr]^{-p}
\end{equation} \end{linenomath}
Die ebenfalls sphärische Dichteverteilung $\rho_b(r)$ des Bulges wird wie von \citet{dehnen98}, allerdings ohne Exzentrizitäten, modelliert:
\begin{linenomath}
\begin{equation}
 \label{equ:bulgerho}
\rho_b(r) = \rho_{0,b} \cdot \Bigl(\frac{r}{r_{0,b}}\Bigr)^{-\gamma_b} ~ \exp\Bigl(\frac{-r^2}{r^2_{t,b}}\Bigr)
\end{equation} \end{linenomath}
Für $r \ll r_{t,b}$ ist damit $\rho_b(r)$ proportional zu $r^{-\gamma_b}$ und nimmt sobald $r \geq r_{t_b}$ stark ab. Die Dichte $\rho_{0,b}$, der Radius $r_{0,b}$ sowie der Exponent $\gamma_b$ sind durch Beobachtungsdaten des \textit{COBE/DIRBE} Satelliten bestimmt worden.

Die Dichte der galaktischen Scheibe (Disk) wird ebenfalls nach einem von \citet{dehnen98} vorgestellten Modell beschrieben. In diesem Modell besteht die galaktische Scheibe aus dem Interstellaren Medium (ISM) sowie der dünnen und der dicken Sternscheibe. Die folgende Gleichung beschreibt die Dichte der Disk:
\begin{linenomath}
\begin{equation}
\label{equ:diskrho}
 \rho_d(r,z) =\rho_{0,d} \cdot \exp\Bigl( - \frac{R_{m,d} - r}{R_d} -\Bigl|  \frac{z}{\sigma_{z,d}}\Bigr| \Bigr)
\end{equation} \end{linenomath}
Dabei sind $R_{m,d}$ und $R_d$ Radien zur Skalierung der Scheibe, die aus Beobachtungsdaten gewonnen wurden. Ebenso sind von \citet{dehnen98} die Ausdehnungen $\sigma_{z,d}$ bezüglich der galaktischen Ebene sowie die Dichten $\rho_{0,d}$ für die drei Komponenten der Scheibe ermittelt worden. 

Die Gesamtdichte $\rho_g$ der Milchstraße berechnet sich aus der Summe der Dichten von Halo, Bulge und Disk:
\begin{linenomath}
\begin{equation}
\label{equ:gesrho}
 \rho_g(r,z) = \rho_h(r) + \rho_b(r) + \rho_d(r,z)
\end{equation} \end{linenomath}

Vereinfachend wird die Flächendichte $\rho_R(r)$ des durch den Einfall von Zwerggalaxien gebildeten Rings aus CDM der Masse $M_R$ über die Breite $R_{max} - R_{min}$ des Rings in diesem Modell als normalverteilt angenommen, der mittlere Radius des Rings $R$ stellt als Erwartungswert der Dichtefunktion den Radius höchster Dichte dar. Die Standardabweichung $\sigma_R$ beschreibt die Größe des Bereichs hoher Dichte im CDM-Ring und berechnet sich über $d = 2 \sigma_R$, wobei $d$ den Durchmesser der verursachenden Zwerggalaxie darstellt.
\begin{linenomath}
\begin{equation}
\rho_R(r) = \frac{M_R}{2 \pi R} ~\cdot~ \frac{1}{\sigma_R \sqrt{2 \pi}} \cdot \exp\Bigl[\frac{(R- r)^2 }{2~\sigma_R^2}  \Bigr]
\end{equation} \end{linenomath}

\begin{figure}[ht]
 \centering
 \includegraphics[width=\columnwidth]{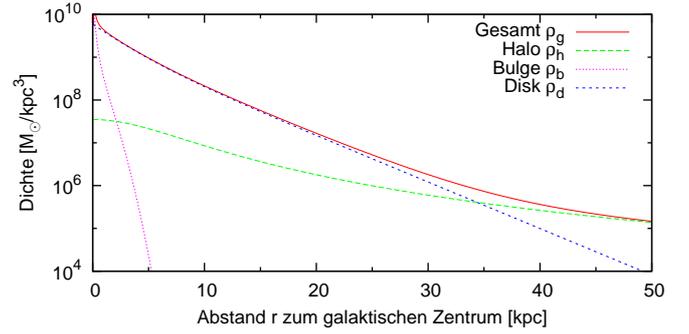}
\caption{Dichteverteilung $\rho(r)$ in der galatischen Ebene der Milchstraße in Abhängigkeit vom Abstand $r$ zum galaktischen Zentrum}
\label{fig:Dichte_in_Milchstr}
\end{figure}

In Tabelle \ref{table:Dichte_in_Milchstr} sind die Parameter der verwendeten Modelle aufgelistet und Abbildung \ref{fig:Dichte_in_Milchstr} zeigt die Dichteverteilung in der Ebene der Milchstraße ($z=0$) in Abhängigkeit vom Abstand $r$ zum galaktischen Zentrum.
\begin{table}[t]
\begin{small}
 \begin{tabular*}{\columnwidth} {@{\extracolsep{\fill}}lllll}
Halo						&	$\rho_{0,h}$ 				& $r_{c,h}$		& $p$ 			&	\\
\hline
\vspace{.03cm}				& $3.50 \cdot 10^{7}$ 			& 8.0 			& 1.5  					& \\

Bulge 					 &	$\rho_{0,b}$ 				& $r_{0,b}$		& $r_{t,b}$ 				&$\gamma_b$\\
\hline
\vspace{.03cm}							& $4.27 \cdot 10^8$ 		& 1.00 				& 1.90  					& 1.80\\

Disk												& $\rho_{0,d}$ 				& $R_{d}$		& $R_{m,d}$ 			& $\sigma_{z,d}$\\
\hline
ISM												& $1.19 \cdot 10^{10}$	&	4.00				&	4.00						&	$4.00 \cdot 10^{-2}$	\\
Dünne											&$7.41 \cdot 10^9$ 		&	2.00				&	0.00						&	$1.80 \cdot 10^{-1}$	\\
Dicke											&$9.53 \cdot 10^7$			&	2.00				&	0.00 					&	$1.00$	
\end{tabular*} 
\end{small}
\caption{Parameter der Dichteverteilungen gemäß \citet{dehnen98} (Disk und Bulge) sowie \citet{saha09} (Halo). Dichten in M$_\odot$/kpc$^3$, Strecken in kpc}
\label{table:Dichte_in_Milchstr}
\end{table}

\subsection{Gravitationspotential}
\label{subsec:Gravitationspotential}
Die Orbits von in die Milchstraße einfallenden Zwerggalaxien hängen von ihrer Gesamtenergie $E_{ges}$ ab, welche sich aus der kinetischen Energie $E_{kin}$ und der potentiellen Energie $E_{pot}$ zusammen setzt. Die potentielle Energie $E_{pot}$ lässt sich aus dem Produkt der Masse der Zwerggalaxie $m$ und dem Gravitationspotential $\Phi(\vect r)$ berechnen. Dabei hängt $\Phi(\vect r)$ vom Positionsvektor $\vect r$ der Zwerggalaxie sowie von der in Abschnitt \ref{subsec:dichte_milchstr} beschriebenen Dichteverteilung $\rho(\vect r)$ ab. Für sphärisch symmetrische Dichteverteilungen wie die des Halos und die des Bulges ist $\Phi_{sph}$ nur von $r=|\vect r|$ abhängig:
{\small
\begin{linenomath}
\begin{equation}
\label{equ:phiSPH}
 \Phi_{sph}(r) = - \int_0^r \frac{G M(s)}{ s{^2} } ds ~~\mbox{mit}~~ M(s) =4 \pi  \int_0^{s} a{^2} \rho(a) da
\end{equation} \end{linenomath}
}
Dabei beschreibt $M(r)$ die Masse innerhalb der Kugel mit dem Radius $r$, welche bei sphärischen Massenverteilungen gemäß dem newtonschen Theorem als Punktmasse bei $r=0$ betrachtet werden kann.

\begin{figure}[ht]
	\centering
	\includegraphics[width=0.45\columnwidth]{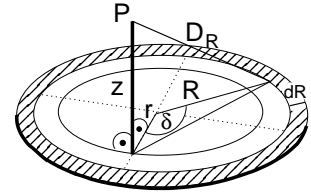}
	\caption{Ring um das galaktische Zentrum mit Radius $R$ und Masse $M_R$ innerhalb der galaktischen Ebene. Der Schnittpunkt der gestrichelten Linien beschreibt die Lage des galaktischen Zentrums.}
	\label{fig:Ring_Pot}
	\end{figure}

Für ringförmige Massenverteilungen innerhalb der galaktischen Ebene um das Zentrum der Milchstraße ist das Gravitationspotential in einem Punkt $P$ abhängig von der Position des Punktes bezüglich der galaktischen Ebene, siehe Abb. \ref{fig:Ring_Pot}.
Der Abstand $D_R$ vom Punkt $P$ zu einem Punkt auf dem Ring mit Radius $R$, abhängig vom Winkel $\delta$, berechnet sich wie folgt:
\begin{linenomath}
\begin{equation}
 D_R(r,z,\delta) = \Bigl( r^2 + z^2 + R^2 - 2Rr\cos(\delta) \Bigr)^{1/2}
\end{equation} \end{linenomath}
Der CDM-Ring wird als Ring der Breite $R_{max} - R_{min}$ innerhalb der galaktischen Ebene modelliert.  Es ergibt sich für das Gravitationspotential des Rings $\Phi_{ring}  = -G dm/D_R$ durch Integration über alle Radien $R$ des Rings und den Winkel $\delta$:
\begin{linenomath}
\begin{equation}
\label{equ:phiRing}
 \Phi_{ring}(r,z) = -G \int_{R_{min}}^{R_{max}}  \int_0^{2\pi} \frac{R ~\rho_R(R) } {D_R(r,z,\delta)} ~d\delta ~ dR
\end{equation} \end{linenomath}
Dabei beschreibt $\rho_R$ die Masse pro Umfang des Rings bei einem bestimmten Radius $R$. 
Um Rechenzeit zu sparen, wird aufgrund der im Vergleich zum CDM-Halo kleinen Masse der Disk die Masse der Disk als Näherung in ihrer gravitativen Auswirkung über das Potential des sphärisch symmetrischen Halos beschrieben.

\subsection{Dynamische Reibung im Halo}
\label{Subsec:DynFriction}
 Bewegt sich eine Zwerggalaxie mit der Geschwindigkeit $v$ durch ein Gebiet der Milchstraße der Dichte $\rho(r)$, so übt diese eine Kraft proportional zu ihrer Masse $m$ auf umliegende Körper, wie Sterne oder Gaswolken, aus. Diese Körper erfahren eine Beschleunigung in Richtung der Zwerggalaxie und führen damit im Bereich in Bewegungsrichtung hinter der Zwerggalaxie  zu einem Dichteanstieg. Durch die Masse in diesem Bereich höherer Dichte wirkt nun eine Kraft auf die Zwerggalaxie, die entgegen ihrer Bewegungsrichtung gerichtet ist. Diese Kraft führt zu einer Abbremsung der Zwerggalaxie,  welche daher Bewegungsenergie und Drehimpuls verliert. Erstmalig beschrieben wurde die dynamische Reibung von \citet{chandrasekhar43}.  Der Prozess ist besonders effektiv bei massiven, langsamen Zwerggalaxien, da diese für eine lange Zeit eine große Kraft auf umliegende Körper ausüben, welche eine große Kraft entgegen der Bewegungsrichtung der Zwerggalaxie bewirken.

In den äußeren Bereichen des sphärischen Halos der Milchstraße können die Neigungen der Orbits von Teilchen im Halo als zufällig verteilt angesehen werden. In diesem Bereich hat sich, aufgrund der zufälligen Orbitneigungen, die Modellierung der Geschwindigkeiten der Teilchen des Halos über eine Maxwellverteilung bewährt \citep{tremaine75}. Unter diesen Annahmen lässt sich mit der folgenden Gleichung die Änderung $\dot v$ pro Zeit der Geschwindigkeit $v$ der Zwerggalaxie bestimmen:
\begin{linenomath}
\begin{equation}
 \label{equ:dotv_dynfrict}
 \dot v = - 4 \pi G^2 m \varrho(R) \ln \Lambda \frac{1}{v^2} \Bigl[ \mbox{erf}\bigl( \frac{v}{v_c} \bigr) - \frac{v}{v_c} \mbox{erf}' \bigl( \frac{v}{v_c} \bigr) \Bigr]
\end{equation} \end{linenomath}
Dabei ist $G$ die Gravitationskonstante und $\ln \Lambda$ ist der Coulomb-Logarithmus, welcher die Reichweite der dynamischen Reibung festlegt. Dieser wird über eine Computersimulation abgeschätzt: Ein Probekörper bewegt sich durch einen das Milchstraßenhalo repräsentierenden Raum mit 1000 Massepunkten, die gravitativ mit dem Probekörper wechselwirken und ihn durch dynamische Reibung abbremsen. Aus der Simulation wird die Geschwindigkeitsänderung pro Zeit $\dot v$ bestimmt und über Gleichung \ref{equ:dotv_dynfrict} auf den entsprechenden Coulomb-Logarithmus $\ln \Lambda$ geschlossen.
Die Simulation führt bei einem Massenfeld der Dichte $\rho_{0,h}$ und einem Probekörper der Masse $m=2.8 \cdot 10^{10}$ M$_\odot$ und der Geschwindigkeit $v_0 = 250$km/s zum Coulomb-Logarithmus $\ln \Lambda = 3.01$. Dieser Wert liegt im Bereich der von \citet{just04} verwendeten Coulomb-Logarithmen für das Milchstraßenhalo.

	\begin{figure}[ht]
	\centering
	\includegraphics[width=0.90\columnwidth]{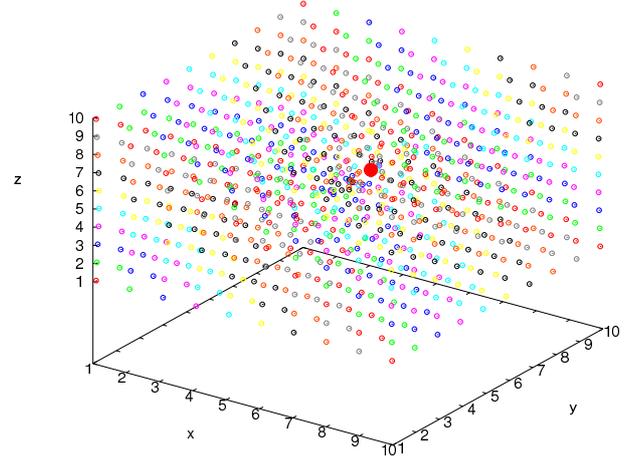}
	\caption{Probekörper (ausgefüllte Kugel) im Massenfeld, wobei die Bewegungsrichtung des Probekörpers der Richtung der positiven y-Achse entspricht. In dem (10kpc)$^3$ großen Volumen befinden sich 1000 Massenpunkte.
}
	\label{fig:nbody}
	\end{figure}

Die Simulationen sind auf numerische Konvergenz getestet. Dazu wurde in einem Simulationslauf die Schrittweite $\Delta t$ zur numerischen Integration der Bewegung der Partikel um Faktor zehn verringert. In einem anderen Simulationslauf wurde die Anzahl der Massenpunkte um Faktor zehn erhöht. Der jeweils ermittelte Coulomb-Logarithmus $\ln \Lambda$ weicht um weniger als $1$ Prozent von $\ln \Lambda = 3.01$ ab.

%
\subsection{Wechselwirkung von Baryonen und Disk}
\label{subsec:disk}
Während die Dunkle Materie beim Einfall der Zwerggalaxie nur über dynamische Reibung Bewegungsenergie verliert, kann die baryonische Materie aus der Zwerggalaxie durch Kollision mit der baryonischen Materie der Milchstraße abgebremst werden. Da Wasserstoff das Element mit der größten Häufigkeit in der Milchstraße ist und stark konzentriert in der Disk vorliegt, werden die Kollisionen als solche zwischen H1-Atomen simuliert. Die Stöße werden als elastisch und zwischen Kugeln des Durchmessers~d mit harten Schalen beschrieben. Ziel ist es in diesem Abschnitt, eine Durchschnittsgeschwindigkeit eines H1-Atoms nach einer Kollision zu berechnen.

Die thermische Bewegung der H1-Atome von Zwerggalaxie und Milchstraßenscheibe führt, durch die geringe Dichte des Wasserstoffgases und damit durch die große freie Weglänge zwischen den H1-Atomen, zu Geschwindigkeitsunterschieden zwischen den einzelnen H1-Atomen. Die Geschwindigkeit eines Wasserstoffatoms in der folgenden Berechnung beschreibt stets die Durchschnittsgeschwindigkeit der Wasserstoffatome von Zwerggalaxie oder Milchstraßenscheibe, die thermischen Geschwindigkeiten mitteln sich damit zu null und werden in der Berechnung nicht beachtet.

\begin{figure}[ht]
	\centering
	\includegraphics[width=0.65\columnwidth]{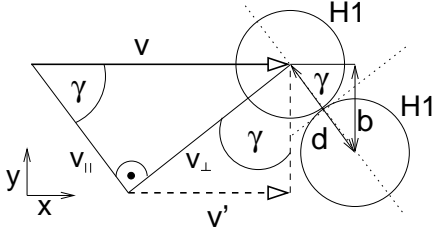}
	\caption{Koordinatensystem, in welchem die Kollision betrachtet wird. Die x-Achse ist so gewählt, dass ihre Richtung jener der Relativgeschwindigkeit $\vect v$ der H1-Atome entspricht. }
	\label{fig:Coll_Coord}
	\end{figure}

In der folgenden Betrachtung wird ein H1-Atom als ruhend angenommen, das andere Atom bewegt sich mit der Relativgeschwindigkeit $\vect v$. Die Richtung der x-Achse des genutzten Koordinatensystems entspricht jener von $\vect v$, siehe Abbildung \ref{fig:Coll_Coord}. 
Entscheidend für die Kollision ist der Kollisionwinkel $\gamma$. Bei einer Kollision verliert das stoßende Atom die Geschwindigkeitskomponente $v_\parallel$ parallel zur Verbindungslinie der beiden H1-Atome an das ruhende Atom und behält selbst lediglich die Geschwindigkeitskomponente $v_\perp$ senkrecht zur Verbindungslinie.

\begin{figure}[ht]
	\centering
	\includegraphics[width=0.45\columnwidth]{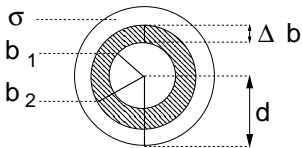}
	\caption{Wirkungsquerschnitt $\sigma$ und Stoßparameterintervall zwischen $b_1$ und $b_2$. Dem Radius $d$ entspricht der Abstand der beiden H1-Atome zum Zeitpunkt der Kollision.}
	\label{fig:bprob}
	\end{figure}

\begin{figure}[ht]
	\centering
	\includegraphics[width=0.40\columnwidth]{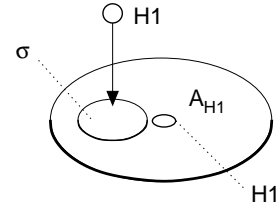}
	\caption{Fläche $A_{H1}$ um ein H1-Atom sowie Wirkungsquerschnitt $\sigma$. Die Wahrscheinlichkeit einer Kollision ist durch das Flächenverhältnis von $\sigma$ und  $A_{H1}$ gegeben.}
	\label{fig:Hsigma}
	\end{figure}

Der Winkel $\gamma$ tritt um die x-Achse des verwendeten Koordinatensystems im Raum rotiert auf. Aus Symmetriegründen bleibt bei Betrachtung aller möglichen Rotationen von $\gamma$ lediglich die x-Komponente $v'$ von $v_\perp$ erhalten, da sich die y-Komponenten bei hinreichend großer Stoßanzahl zu null mitteln. Ist der Winkel $\gamma$ bekannt, so ergibt sich für $v'$ (vergleiche Abbildung \ref{fig:Coll_Coord}):
\begin{linenomath}
\begin{equation}
 v'(\gamma) = v \sin^2 \gamma
\end{equation} \end{linenomath}
Im nächsten Schritt wird die Wahrscheinlichkeit eines Stoßwinkels $\gamma$ berechnet. Die Wahrscheinlichkeit einer Kollision hängt von der Größe des Stoßparameters $b$ ab: Kleine Stoßparameter, die zu frontalen Kollisionen führen, haben kleine Wahrscheinlichkeiten; große Stoßparameter, die lediglich das gegenseitige Streifen der H1-Atome bewirken, haben höhere Wahrscheinlichkeiten. Abbildung \ref{fig:bprob} zeigt ein Stoßparameterintervall von $b_1$ bis $b_2$. Die Wahrscheinlichkeit dieses Stoßparameterintervalls entspricht seinem Flächenverhältnis zum Wirkungsquerschnitt $\sigma$.
Bei bekanntem Wirkungsquerschnitt $\sigma = \pi d^2$ gilt für die kumulierte Wahrscheinlichkeit $p_{cum}(b)$ eines Stoßparametes $b$ daher:
\begin{linenomath}
\begin{equation}
\label{equ:PcumStossparam}
 p_{cum}(b) = \Bigl[ \frac{\pi b^2}{\sigma} \Bigr]^{b}_{0}
\end{equation} \end{linenomath}
Aus Abbildung \ref{fig:Coll_Coord} ist zu erkennen, dass $b = d \sin \gamma$. Setzt man dies in Gleichung \ref{equ:PcumStossparam} ein und leitet ab, um zur partiellen Wahrscheinlichkeit eines Winkels $\gamma$ zu gelangen, erhält man:
\begin{linenomath}
\begin{equation}
 p_{par}(\gamma) = \frac{d \sin^2 \gamma}{d\gamma} = 2 \sin \gamma \cos \gamma
\end{equation} \end{linenomath}
Da nun die Wahrscheinlichkeit eines Stoßwinkels $\gamma$ sowie die mit einem bestimmten $\gamma$ resultierende Geschwindigkeit $v'$ bekannt ist, lässt sich über Gewichtung von $v'(\gamma)$ mit $p_{par}(\gamma)$ und Integration über alle möglichen $\gamma$ auf die durchschnittliche Geschwindigkeit $\bar v'$ nach einem Stoß schließen:
\begin{linenomath}
\begin{equation}
 \bar v' = 2 v \int_0^{\frac{\pi}{2}} \sin^3 \gamma \cos \gamma d\gamma = \frac{v}{2}
\end{equation} \end{linenomath}
Für die Geschwindigkeit $v_{trans}$ eines H1-Atoms des Wasserstoffgases der Zwerggalaxie nach Transit der Disk der Milchstraße muss die Flächendichte $\bar \rho_{H1}$ an H1-Atomen in der Disk bekannt sein. Aus $1/\bar \rho_{H1}$ ergibt sich die durchschnittliche jedes H1-Atom umgebende Fläche $A_{H1}$. Das Verhältnis von $\sigma$ und dieser Fläche $A_{H1}$ beschreibt die Wahrscheinlichkeit für eine Kollision, siehe Abbildung \ref{fig:Hsigma}.
Es ergibt sich aus diesen Überlegungen die Geschwindigkeit $v_{trans}$ eines H1-Atoms nach Transit der galaktischen Scheibe über:
\begin{linenomath}
\begin{equation}
 v_{trans} = v \Bigl(\frac{1}{2}\Bigr)^\frac{\sigma}{A_{H1}}
\end{equation} \end{linenomath}
Dabei ist $v$ die Relativgeschwindigkeit des H1-Atoms der Zwerggalaxie zu den H1-Atomen der Disk bei Eintritt in die Disk.

%

\subsection{Gezeitenkräfte}
\label{subsec:TidalDisruption}
Fällt eine Zwerggalaxie in die Milchstraße ein, so steigen mit abnehmendem Abstand zum galaktischen Zentrum die auf die Zwerggalaxie wirkenden Gezeitenkräfte. Ist die Differenz $F_{M,1} - F_{M,2} = \Delta F_M$ der auf die Zwerggalaxie wirkenden Kräfte im Gravitationspotential der Milchstraße größer als jene die Zwerggalaxie zusammen haltende Kraft $F_{zw}$, wird die Zwerggalaxie zu einem Sternstrom zerrissen. 
	\begin{figure}[ht]
	\centering
	\includegraphics[width=0.55\columnwidth]{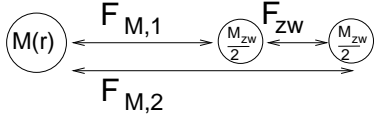}
	\caption{Kräfte zwischen zwei Massen $M_{zw}/2$ im Gravitationspotential der Milchstraße}
	\label{fig:tidal}
	\end{figure}
Die Kraft $F_{zw}$ wird vereinfachend als jene Gravitationskraft beschrieben, die zwischen zwei Körpern der Masse $M_{zw}/2$ im Abstand $d$ wirkt, wobei $M_{zw}$ die Masse und $d$ den Durchmesser der Zwerggalaxie darstellt. Um $\Delta F_M$ zu berechnen, wird die Differenz der Kräfte zwischen der Milchstraße und der Masse $M_{zw}/2$ in einem Abstand $R$ und $R+d$ zum galaktischen Zentrum bestimmt. Dabei ist $R$ der Abstand des galaktischen Zentrums zur dem galaktischen Zentrum zugewandten Seite der Zwerggalaxie. Es folgt nach einigem Umformen, dass die Zwerggalaxie stabil ist, so lange gilt:
\begin{linenomath}
\begin{equation}
\label{equ:Tidal}
 \frac{M_{zw}}{2 d^2} > \frac{M(r)}{r^2} - \frac{M(r+d)}{(r+d)^2}
\end{equation} \end{linenomath}
Dabei ist $M(r)$ die Masse der Milchstraße innerhalb des Radius $r$. Für Radien kleiner als $r_t$ ist die oben genannte Bedingung nicht mehr erfüllt und die Zwerggalaxie wird  zu einem Sternstrom zerrissen.

%

\subsection{Numerische Integration}
\label{subsec:numint}
Die Bewegung der Zwerggalaxie wird durch das Gravitationspotential $\Phi(\vect r)$ der Milchstraße, siehe Abschnitt \ref{subsec:Gravitationspotential}, bestimmt. Die von der Zwerggalaxie erfahrene Beschleunigung $\vect{\ddot r} $ ergibt sich aus der Ableitung des Gravitationspotentials nach dem Positionsvekor und führt zur folgenden Differentialgleichung 2. Ordnung:
\begin{linenomath}
\begin{equation}
 \vect {\ddot  r} = -\frac{d\Phi(\vect r)}{d\vect r}
\end{equation} \end{linenomath}
 Diese wird durch zweifaches Anwenden des vierstufigen Runge Kutta-Verfahrens (siehe zum Beispiel  \citealp{hellings94}) numerisch integriert.  Um aus der Beschleunigung $ \vect{\ddot  r}$ auf die Geschwindigkeit $\vect{\dot  r}$ zu schließen, werden vier Steigungen $k_{\vect{\dot r}}$ benötigt:
{\small\begin{eqnarray}
 k_{1 \vect{\dot r}} &=&  -\frac{d\Phi}{d\vect r} \Bigl( \vect{r} \Bigr) \\
 k_{2 \vect{\dot r}} &=&  -\frac{d\Phi}{d\vect r} \Bigl( \vect{r} +  k_{1 \vect{ r}} \frac{\Delta t}{2} \Bigr) \\
 k_{3 \vect{\dot r}} &=&  -\frac{d\Phi}{d\vect r} \Bigl( \vect{r} +  k_{2 \vect{ r}} \frac{\Delta t}{2} \Bigr) \\
 k_{4 \vect{\dot r}} &=&  -\frac{d\Phi}{d\vect r} \Bigl( \vect{r} +  k_{3 \vect{ r}} \Delta t \Bigr)
\end{eqnarray}}
Dabei ist $\Delta t$ die für die Genauigkeit der Integration entscheidende Schrittweite. Die Geschwindigkeit $\vect{\dot r}$ ergibt sich aus dem gewichteten Mittel der 4 Steigungen nach Multiplikation mit der Schrittweite:
\begin{linenomath}
\begin{equation}
 \vect{\dot{r}}_{t+1} =  \vect{\dot{r}}_{t} + \frac{k_{1 \vect{\dot r}}  +2k_{2 \vect{\dot r}} +2k_{3 \vect{\dot r}} + k_{4 \vect{\dot r}}}{6} \Delta t
\end{equation} \end{linenomath}
Die Steigungen $k_{\vect {\dot r}}$ müssen abwechselnd mit den Steigungen $k_{\vect {r}}$ berechnet werden:
{\small\begin{eqnarray}
 k_{1 \vect{r}} &=&  \vect{\dot{r}} \\
 k_{2 \vect{r}} &=&  \vect{\dot r} +  k_{1 \vect{\dot r}} \frac{\Delta t}{2}  \\
 k_{3 \vect{r}} &=&  \vect{\dot r} +  k_{2 \vect{\dot r}} \frac{\Delta t}{2}  \\
 k_{4 \vect{r}} &=&  \vect{\dot r} +  k_{3 \vect{\dot r}} \Delta t 
\end{eqnarray}}
Der neue Positionsvektor $\vect r$ ergibt sich aus:
\begin{linenomath}
\begin{equation}
 {\vect r}_{t+1} = {\vect r}_{t} + \frac{k_{1 \vect{r}}  +  2k_{2 \vect{r}} + 2k_{3 \vect{r}} +  k_{4 \vect{r}}} {6} \Delta t
\end{equation} \end{linenomath}
Da $\vect r$ eine vektorielle Größe ist, muss die eben beschriebene Zeitintegration für jede Komponente einzeln ausgeführt werden.


\section{Ergebnisse}
In diesem Abschnitt werden die Ergebnisse der in Abschnitt \ref{sec:sim} beschriebenen Simulation dargestellt.

Das sphärisch symmetrische Dichteprofil von Halo $\rho_h(r)$ und Bulge $\rho_b(r)$ wird mit einer räumlichen Auflösung von $1/30$kpc diskretisiert. Mit Gleichung \ref{equ:phiSPH} und numerischer Berechnung von $M(r)$ wird damit das sphärisch symmetrische Gravitationspotential beschrieben, aus Gleichung \ref{equ:phiRing} wird auf das Gravitationspotential des Rings aus CDM geschlossen. 

 Die Schrittweite $\Delta t$ der Zeitintegration beträgt $1 \cdot 10^6$ Jahre. Es wird ein Zeitraum dem Alter des Universums entsprechend von $13.7 \cdot 10^9$ Jahren  simuliert \citep{spergel03}.

\label{sec:ergebnisse}
	\begin{figure*}[ht]
	\centering
	\includegraphics[width=1.6\columnwidth]{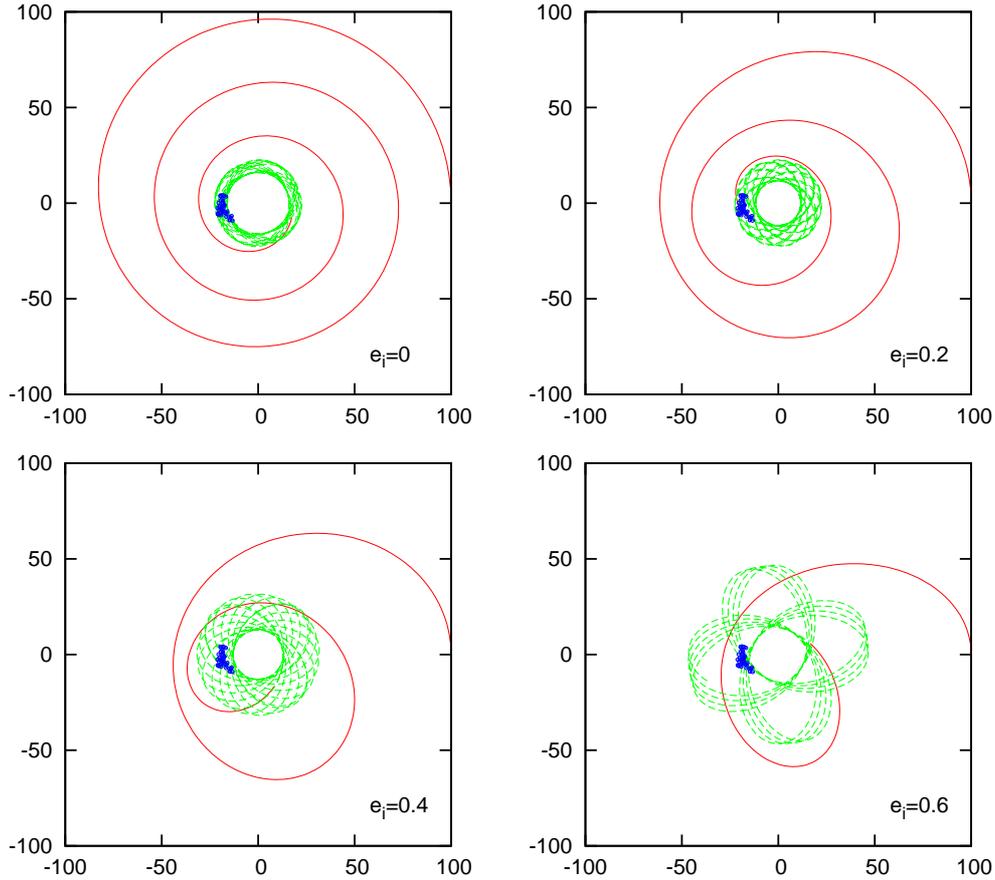}
	\caption{Orbits einer Zwerggalaxie mit der Masse $m = 2.8\cdot10^{10} \mbox{M}_\odot$ im Gravitationspotential der Milchstraße. Es sind Orbits mit  verschiedenen Exzentrizitäten $e_i$ zu Beginn der Simulation dargestellt. Die Inklination ist bei allen Orbits mit $i=0$ gewählt. Die Galaxie startet beim Radius $r_i = 100$ kpc und bewegt sich in Richtung des galaktischen Zentrums (durchgängige Linie). Bei Erreichen von $r_t$ wird sie durch  Gezeitenkräfte zerrissen. Das Orbit ist von dort an mit gestrichelter Linie weitergeführt, entlang dieser Bahn werden sich Sterne, Wasserstoffgas und CDM der Zwerggalaxie ansammeln. Die Punkte markieren die von \citet{crane03} beobachteten Sterne der Ringstruktur.}
	\label{fig:Orbits_Einzeln}
	\end{figure*}

\subsection{Exzentrizität des Orbits}
\label{SubSec:Erg:ECC}
Die bei 14 kpc Entfernung zum galaktischen Zentrum beobachtete Ansammlung von kalter Dunkler Materie, Sternen und Wasserstoffgas wird als annähernd ringförmig angenommen \citep{kalberla07, ibata03}, somit kann von einer kleinen numerischen Exzentrizität $e$ des Rings ausgegangen werden. 

Es soll nun geklärt werden, in wie weit die Exzentrizität des Orbits der Zwerggalaxie sich auf den entstehenden Ring auswirkt. Die Exzentrizität des Orbits der Zwerggalaxie lässt sich, da aufgrund der wirksamen dynamischen Reibung und der Massenverteilung $M(r) \neq \mbox{const}$ die Zwerggalaxie keine reinen Keplerellipsen beschreibt, nicht über die herkömmliche numerische Exzentrizität einer Ellipse darstellen. In Anlehnung an die Exzentrizität einer Ellipse $e$ wird hier die Exzentrizität $\varepsilon_t$ wie folgt definiert:
\begin{linenomath}
\begin{equation}
 \varepsilon_t := \frac{r_{t,max} - r_{t,min}}{r_{t,max} + r_{t,min}}
\end{equation} \end{linenomath}
Dabei entsprechen die Radien $r_{t,max}$ und $r_{t,min}$ dem jeweils zeitlich letzten maximalen und minimalen Abstand der Zwerggalaxie zum galaktischen Zentrum. 

\begin{figure}[t]
	\centering
	\includegraphics[width=\columnwidth]{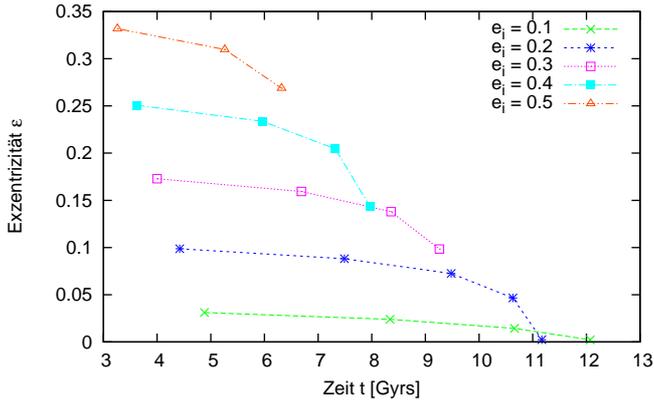}
	\caption{Entwicklung der Exzentrizität $\varepsilon$ des Orbits einer Zwerggalaxie in Abhängigkeit der Zeit $t$. Der Zusammenhang ist für unterschiedliche Werte der numerischen Exzentrizität $e_i$ am Anfang der Simulation dargestellt. Das Ende einer Linie entspricht dem Zeitpunkt der Zerstörung der Galaxie durch Gezeitenkräfte, nach welchem $\varepsilon_t$ nahezu konstant bleibt.}
	\label{fig:EntwicklungEcc}
	\end{figure}

 Um das Verhalten von $\varepsilon_t$ in Abhängigkeit der Zeit $t$ zu untersuchen, wird eine Zwerggalaxie beim Radius $r_i = 100$ kpc mit der Geschwindigkeit $v_i$ gestartet und bei jedem Erreichen des Perizenters die jeweilige Exzentrizität $\varepsilon_t$ bestimmt.
Dabei stellt $r_i$ den größten Abstand zum galaktischen Zentrum dar. Für diesen Punkt wird für verschiedene numerische Exzentrizitäten $e_i$ einer Keplerellipse die jeweilige Apoapsisgeschwindigkeit berechnet (siehe dazu zum Beispiel \citealp{karttunen07}), welche als Anfangsgeschwindigkeit $v_i = v_c \sqrt{1-e_i}$ der Zwerggalaxie verwendet wird. Dabei ist $v_c$ die stabile Kreisbahngeschwindigkeit für $r_i$.
Abbildung \ref{fig:EntwicklungEcc} zeigt die Entwicklung der Exzentrizität $\varepsilon$ in Abhängigkeit der Zeit $t$ für verschiedene numerische Exzentrizitäten $e_i$, die zum Berechnen der Startgeschwindigkeit verwendet wurden.

Nach der Zerstörung der Zwerggalaxie durch Gezeitenkräfte liegt diese als Sternstrom vor und erfährt aus diesem Grund weniger dynamische Reibung. Dieser Sternstrom behält daher die Exzentrizität des Orbits der ihn verursachenden Zwerggalaxie weitgehend bei und führt auch zu einem Ring entsprechend der Exzentrizität $\varepsilon$ des Orbits der Zwerggalaxie.

Die Exzentrizität $\varepsilon$ des Orbits nimmt im Laufe der Zeit durch Effekte dynamischer Reibung ab, der Orbit wird also kreisförmiger. Das ist ein häufig beobachteter Prozess bei Simulationen mit konstantem Coulomb Logarithmus $\ln \Lambda$ \citep{hashimoto03}. Dieser ist jedoch zu langsam, um Orbits mit numerischen Exzentrizitäten $e_i > 0.2$ so weit kreisförmig zu machen, als dass die Masse der Zwerggalaxie in der beobachteten Weise um das galaktische Zentrum verteilt würde, siehe Abbildung \ref{fig:Orbits_Einzeln}. Daher kann die anfängliche numerische Exzentrizität $e_i$ des Orbits einer eingefallenen und an der Ringenstehung beteiligten Zwerggalaxie nicht größer als etwa 0.2 gewesen sein. 

\subsection{Inklination des Orbits}
\begin{figure}[t]
	\centering
	\includegraphics[width=\columnwidth]{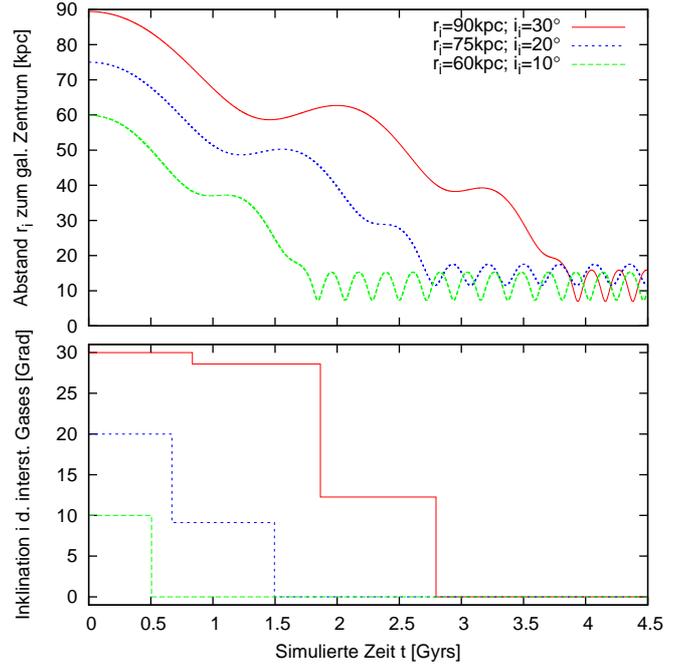}
	\caption{Zeitliche Entwicklung des Abstands $r_i$ der Zwerggalaxie zum galaktischen Zentrum (oberes Diagramm) sowie der Inklination $i$ des interstellaren Wasserstoffgases der Zwerggalaxie beim mehrmaligen Durchkreuzen der galaktischen Scheibe (unteres Diagramm).}
	\label{fig:EntwicklungIncl}
	\end{figure}

	Kreuzt die Zwerggalaxie auf ihrem Orbit die galaktische Scheibe der Milchstraße, kommt es zu Kollisionen zwischen dem interstellaren Wasserstoffgas der Milchstraßenscheibe und dem Wasserstoffgas der Zwerggalaxie, siehe Abschnitt \ref{subsec:disk}.
	Je dichter die Zwerggalaxie dabei dem galaktischen Zentrum ist, desto höher ist die Dichte des Wasserstoffgases an der Stelle, an welcher die Zwerggalaxie die Milchstraßenscheibe durchkreuzt, siehe Gleichung \ref{equ:diskrho}. Damit steigt mit sinkendem Abstand zum galaktischen Zentrum die Wahrscheinlichkeit einer Kollision zwischen Wasserstoff aus der Milchstraße und Wasserstoff aus der Zwerggalaxie.
	Abbildung \ref{fig:EntwicklungIncl} zeigt für drei Zerggalaxien der Masse $2.8 \cdot 10^{10} M_\odot$ die Entwicklung der Inklination $i$ im Laufe der Simulation für verschiedene anfängliche Inklinationen $i_i$ und verschiedene Startradien $r_i$ des Orbits, die anfängliche Exzentrizität $e_i$ beträgt $0.1$.
	Die Abbildung zeigt, dass bereits nach wenigen Umläufen um das galaktische Zentrum die Bewegungsenergie der Geschwindigkeitskomponente der Zwerggalaxie senkrecht zur galaktischen Scheibe vom Wasserstoffgas der Milchstraße aufgenommen wurde. Dadurch sinkt die Inklination des Wasserstoffgases der Zwerggalaxie schnell. Die vom Wasserstoffgas der Milchstraße aufgenommene Bewegungsenergie kann durch weitere Kollisionen innerhalb der galaktischen Scheibe in Wärmeenergie umgewandelt werden. 

	Die Dunkle Materie wechselwirkt nicht elektromagnetisch, siehe Abschnitt \ref{sec:einleitung}. Deshalb kann die Inklination der Dunklen Materie der Zwerggalaxie nicht wie die Inklination des Wasserstoffgases durch Kollision abnehmen. Durch dynamische Reibung in abgeflachten CDM-Halos ist ein Prozess bekannt, welcher beschreibt, wie sich rein gravitativ beeinflusst die Inklination des Orbits einer Zwerggalaxie ändern kann \citep{penarrubia04}. Dieser findet in dem hier verwendeten Modell jedoch keine Anwendung, da das simulierte CDM-Halo sphärisch symmetrisch ist, siehe Gleichung \ref{equ:pseudoisothermal}.
	Fällt eine Zwerggalaxie in die Milchstraße mit einem Orbit deutlich außerhalb der galaktischen Ebene ein, kommt es zu einer Trennung von Wasserstoffgas und Dunkler Materie der Zwerggalaxie, da das Wasserstoffgas Bewegungsenergie an die Milchstraßenscheibe abgeben und dabei die Inklination verringern kann, die Dunkle Materie hingegen nicht.	

\subsection{Masse der Galaxie}
\label{SubSec:GalaxMasse}
Damit eine Zwerggalaxie vom Startradius $r_i$ bis zum Radius $r_t$, an welchem sie durch Gezeitenkräfte zerrissen wird, gelangt, muss sie über dynamische Reibung (siehe Abschnitt \ref{Subsec:DynFriction}) Bewegungsenergie abgeben. Die in der simulierten Zeit über dynamische Reibung maximal abzugebene Bewegungsenergie hängt von der Masse der Zwerggalaxie, dem Startradius $r_i$ sowie von der ursprünglichen Exzentrizität $e_i$ ihres Orbits ab. 

\begin{figure}[ht]
	\centering
	\includegraphics[width=\columnwidth]{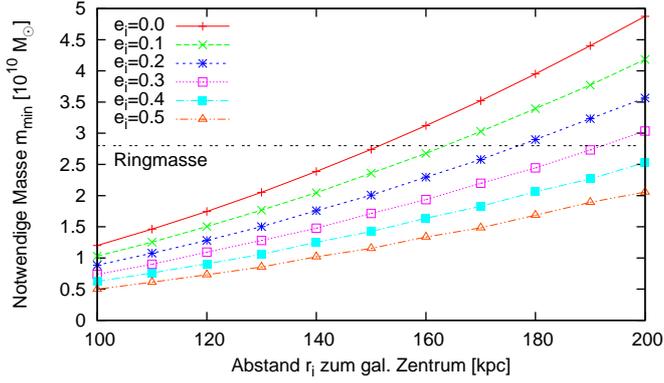}
	\caption{Mindestmasse $m_{min}$ einer Zwerggalaxie, um vom Startradius $r_i$ über dynamische Reibung innerhalb der simulierten Zeit $r_t$ zu erreichen. Der Zusammenhang von $m_{min}$ und $r_i$ ist für verschiedene Exzentrizitäten $e_i$ dargestellt.}
	\label{fig:MasseVsRadius}
	\end{figure}

Die Simulation wird in dieser Untersuchung genutzt, um die minimal nötige Masse $m_{min}$ der Zwerggalaxie zu bestimmen, welche erforderlich ist, um sie vom Startradius $r_i$ über dynamische Reibung auf den Radius $r_t = 14$ kpc innerhalb der simulierten Zeit von $13.7\cdot10^9$ Jahren zu bringen. Die ursprüngliche Exzentrizität $e_i$ des Orbits muss berücksichtigt werden, da mit steigender Exzentrizität $e_i$ die Zwerggalaxie auf ihrem Orbit in Bereiche höherer Dichte gelangt, welche die dynamische Reibung begünstigen. Dies ist auch aus Abbildung \ref{fig:Orbits_Einzeln} ersichtlich: Je größer $e_i$, desto dichter kommt die Zwerggalaxie auf ihrem Orbit dem galaktischen Zentrum und daher den Bereichen hoher Dichte. In Abbildung \ref{fig:MasseVsRadius} ist die nötige Mindestmasse $m_{min}$ in Abhängigkeit des Startradius $r_i$ für verschiedene $e_i$ aufgetragen.

Mit dem Startradius $r_i$ steigt die Mindestmasse $m_{min}$. Bei $e_i = 0$ ist zu erkennen, dass der maximale Startradius $r_i$ im Bereich von 150 kpc liegt, da für größere Radien die Masse der Zwerggalaxie die des beobachteten Rings dunkler Materie überstiege.
Bei kleinen Startradien und höheren Exzentrizitäten $e_i$ reichen allerdings bereits Zwerggalaxien mit Massen um Faktor 5 kleiner als der Ringmasse, um bis zu $r_t$ innerhalb der simulierten Zeit zu gelangen.

\subsection{Einfall mehrerer Galaxien}
\label{SubSec:Erg:MehrereGalax}
Wie in Abschnitt \ref{SubSec:GalaxMasse} gezeigt, können auch Zwerggalaxien mit deutlich kleineren Massen als der Ringmasse bis zum Abstand von 14 kpc zum galaktischen Zentrum durch dynamische Reibung abgebremst werden.
Dies ermöglicht die Vermutung, dass der Ring nicht aus einer großen, sondern aus mehreren kleinen Zwerggalaxien entstanden sein könnte. An einem Beispiel soll nun konkret gezeigt werden, dass der Einfall mehrerer Zwerggalaxien ebenfalls zu einer ringförmigen Ansammlung kalter Dunkler Materie bei $r=14$ kpc führen kann.

Dabei wird der Einfall von drei Zwerggalaxien G$_n$ mit Massen $m$ zwischen $6\cdot10^{9}\mbox{M}_\odot$ und $1.3\cdot10^{10}\mbox{M}_\odot$ simuliert. Die Parameter der Galaxien sind in Tabelle \ref{table:Parameter_SimVieleGal} aufgelistet.
\begin{table}[t]
\begin{small}
 \begin{tabular*}{\columnwidth} {@{\extracolsep{\fill}}llllll}
Galaxie								& $m$ 							& $d$				& $r_i$ 			& $i_i$	&$e_i$\\
\hline
G$_1$								& $9 \cdot 10^{9}$	&	4.2				&	80.0				&	14	& 0	\\
G$_2$								& $1.3 \cdot 10^{10}$	&	5.0				&	70.0				&	12 	& 0.1	\\
G$_3$								& $6 \cdot 10^{9}$	&	3.6				&	60.0 			&	10 	& 0	\\
\end{tabular*} 
\end{small}
\caption{Parameter der Simulation des Einfalls mehrerer Galaxien. Masse $m$ in M$_\odot$, Durchmesser $d$ und Startradius $r_i$ in kpc, Inklination $i_i$ in Grad und numerische Exzentrizität $e_i$. Die Durchmesser $d$ werden so gewählt, dass die Zwerggalaxie um den Radius $r_t\approx14kpc$ durch Gezeitenkräfte zerrissen wird, siehe Gleichung \ref{equ:Tidal}. Diese liegen im Bereich der für Zwerggalaxien dieser Größenordnung verwendeten und durch Beobachtung bestätigten Durchmesser, siehe zum Beispiel \citet{bekki08}.}
\label{table:Parameter_SimVieleGal}
\end{table}
Bei mehreren einfallenden Zwerggalaxien erhöhen jene, die den jeweiligen Abstand zum galaktischen Zentrum $r_t$ erreichen, an welchem sie durch Gezeitenkräfte zerrissen werden, die relative Dichte der Milchstraße um diesen Radius (siehe Abschnitt \ref{subsec:TidalDisruption}). Durch die gestiegene Dichte bei $r_t$ wird das Zerreißen einer weiteren einfallenden Zwerggalaxie an diesem Radius begünstigt. Dieser Prozess erschwert, dass Zwerggalaxien geringer Dichte weiter als bis zum Radius $r_t$ in die Milchstraße hinein geraten und trägt dabei sich selbst mit jeder zerrissenen Zwerggalaxie verstärkend zu einer Ringbildung um den Radius $r_t$ bei. Damit wird deutlich, dass der Einfall mehrerer Zwerggalaxien nicht zwingend mehrere Ringe mit sich bringt sondern, wie beobachtet, das Zerreißen der Zwerggalaxien an einem bestimmten Radius und damit die Bildung eines einzelnen Rings begünstigt wird. 

\subsection{Rotationskurve}

	\begin{figure*}[ht]
	\centering
	\includegraphics[width=2\columnwidth]{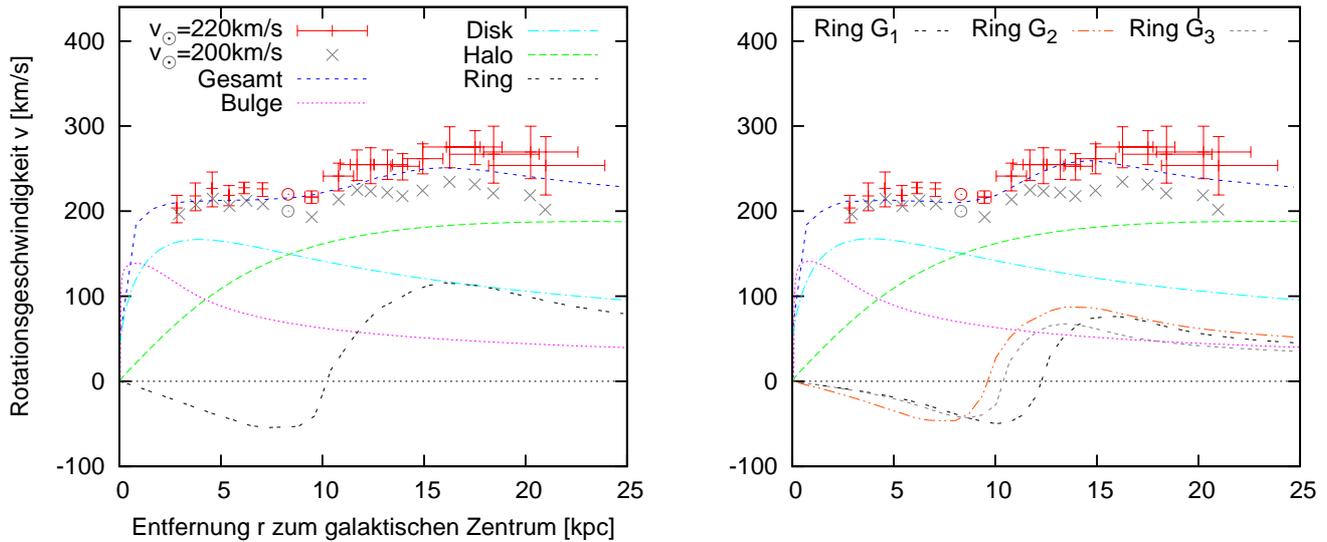}
	\caption{Rotationskurve der Milchstraße mit Ring dunkler Materie als Ergebnis des Einfalls von einer (links) und drei (rechts) Zwerggalaxien. Die Beiträge der einzelnen Komponenten der Milchstraße zur Gesamtgeschwindigkeit sind dargestellt. Als Vergleichswerte (Messreihen mit $v_{\odot}=220$km/s und $v_{\odot}=200$km/s) sind von \citet{honma97} zusammengestellte Messwerte eingetragen. Beide Rotationskurven liegen im Wesentlichen innerhalb der Fehlerbalken dieser Messwerte. }
	\label{fig:Rotationskurven}
	\end{figure*}

In Abbildung \ref{fig:Rotationskurven} sind zwei Rotationskurven dargestellt. Die linke Rotationskurve ergibt sich aus der Simulation des Einfalls nur einer den Ring kalter Dunkler Materie verursachenden Zwerggalaxie mit der Masse $m=2.8\cdot10^{10}$M$_{\odot}$, dem Durchmesser $d=6.48$kpc und einem Orbit mit $e_i=0$ und $i_i=0$. Die rechte Rotationskurve stellt die Rotationskurve der Simulation des Einfalls dreier Zwerggalaxien dar, deren Eigenschaften in Tabelle \ref{table:Parameter_SimVieleGal} aufgelistet sind. 
Als Vergleichswerte sind von \citet{honma97} zusammengestellte Messwerte für $v_{\odot}=220$km/s und $v_{\odot}=200$km/s eingetragen, wobei $v_{\odot}$ die Rotationsgeschwindigkeit der Sonne um das galaktische Zentrum bei $r_{\odot} = 8.3$kpc beschreibt. Die Fehlerbalken der Messung mit $v_{\odot}=200$km/s entsprechen im Wesentlichen denen der Messwerte mit $v_{\odot}=220$km/s und sind der Übersicht halber nicht mit eingezeichnet.

Befindet sich ein Körper zwischen dem CDM-Ring und dem galaktischen Zentrum, wirkt auf diesen eine Kraft in Richtung des galaktischen Zentrums, welche sich aus dem Gravitationspotential der sphärisch symmerischen Dichteverteilung berechnen lässt, siehe Gleichung \ref{equ:phiSPH}. In entgegengesetzte Richtung wirkt auf den Körper eine Kraft, welche sich aus dem Gravitationspotential des CDM-Rings ergibt, siehe Gleichung \ref{equ:phiRing}. Deshalb können Körper zwischen Ring und galaktischem Zentrum langsamer rotieren als sie es ohne den Ring könnten. Dieser Einfluss des Rings ist in den negativen Geschwindigkeiten in Abbildung \ref{fig:Rotationskurven} ausgedrückt.

Sowohl die Rotationskurve gemäß dem Modell des Einzeleinfalls einer Zwerggalaxie, als auch jene nach dem Modell des Einfalls mehrerer Zwerggalaxien liegen innerhalb der Fehlerbalken der von \citet{honma97} zusammengestellten Messwerte der Rotationsgeschwindigkeit. Beide Modelle können den leichten Einbruch der Rotationsgeschwindigkeit bei $r\approx$10kpc reproduzieren, steigen jedoch darauf etwas flacher an als die gemessenen Rotationsgeschwindigkeiten. Auch für große $r$ liegen die Simulationsergebnisse noch innerhalb der Fehlerbalken. Letztere sind jedoch deutlich größer als im Bereich kleiner Abstände zum galaktischen Zentrum.

\section{Diskussion}
An dieser Stelle werden die in Abschnitt \ref{sec:ergebnisse} vorgestellten Ergebnisse der Simulation auf Bedeutung und Aussagekraft hin überprüft. Dazu werden zuerst die Aussagekraft der Ergebnisse einschränkende Fehlerquellen bewertet.

\subsection{Fehlerquellen}
Die Ergebnisse entstammen einem semianalytischen Modell des Einfalls einer und mehrerer Zwerggalaxien in die Milchstraße. Bezeichnend für ein semianalytisches Modell ist, dass sich die Ergebnisse lediglich aus den explizit modellierten Prozessen und derer Zusammenwirkung ergeben. Wichtig für aussagekräftige Ergebnisse ist also eine umfassende Vorstellung aller für das zu simulierende System entscheidenden Prozesse. Modelliert wurde die gravitative Wechselwirkung zwischen der Zwerggalaxie und der Milchstraße mit durch Beobachtungsdaten gestützten Dichteverteilungen der Milchstraße \citep{navarro96, dehnen98}. 

Das verwendete Gravitationspotential (siehe Abschnitt \ref{subsec:Gravitationspotential}) stellt zwar eine Näherungslösung dar, führt allerdings zu Rotationskurven innerhalb der Fehlerbalken von Messungen der Rotationsgeschwindigkeit \citep{honma97}. Dies spricht für die Zulässigkeit der Näherungen.

Das Modell zum Zerreißen der Zwerggalaxie durch Gezeitenkräfte berücksichtigt weder den Massenverlust der Zwerggalaxie auf ihrem Orbit, noch beschreibt es den Prozess des Zerreißens. Das Berechnen des Abstands $r_t$ zum galaktischen Zentrum, an welchem die Zwerggalaxie durch Gezeitenkräfte zerrissen wird, ist jedoch am Zerreißen der großen Magellanschen Wolke getestet. Der von \citet{tremaine75} publizierte Radius $r_t$ ist knapp Faktor 2 größer als der selbst berechnete, allerdings berücksichtigt das hier verwendete Modell im Gegensatz zu dem von \citet{tremaine75} verwendeten die Auswirkungen des Gravitationspotentials der Disk.

Die Reichweite der dynamische Reibung wird über den Coulomb-Logarithmus $\ln \Lambda$ festgelegt. Dieser wurde mit Hilfe einer Computersimulation abgeschätzt, siehe Abschnitt \ref{Subsec:DynFriction}. Je nach Geometrie des Halos und Position der Zwerggalaxie im Halo ändert sich mit dem für die dynamische Reibung wirksamen Bereichs auch $\ln \Lambda$, in diesem Modell wird jedoch vereinfachend ein konstanter Coulomb-Logarithmus verwendet. Dieser konstante $\ln \Lambda$ liegt im Bereich der von \citet{just04} publizierten Coulomb-Logarithmen für das Milchstraßenhalo.

Die durch Kollision abgebremste baryonische Materie der Zwerggalaxie befindet sich gemeinsam mit der CDM im Potential der Zwerggalaxie und führt daher durch gravitative Wechselwirkung auch zu einer Abbremsung der CDM. Allerdings ist die Masse der baryonischen Materie klein im Vergleich zu der Masse der CDM, weshalb diese Form der Abbremsung der CDM nicht berücksichtigt wurde.

Das Verfahren zur numerischen Zeitintegration des Orbits  wurde auf numerische Konvergenz getestet. Dazu wurden Orbits mit einer zwei Größenordnungen kleineren Schrittweite $\Delta t$ als der sonst verwendeten simuliert. Es ließen sich numerisch keine Unterschiede feststellen.

\subsection{Bedeutung der Ergebnisse}
Die Simulationen können für physikalische Eigenschaften der Zwerggalaxie und ihres Orbits Ausschlussgrenzen setzen.  Die maximale anfängliche Exzentrizität $e_i$ des Orbits der Zwerggalaxie kann auf einen Wert von $\approx 0.2$ begrenzt werden, siehe Abschnitt \ref{SubSec:Erg:ECC}. 

Die notwendige Masse der Zwerggalaxie, um vom Startradius $r_i$ bis zum Ringradius bei $R=14$ kpc über dynamische Reibung zu gelangen, ist für $r_i < 150$  kpc deutlich geringer als die Ringmasse, siehe Abbildung \ref{fig:MasseVsRadius}. Dies führt zur Annahme, dass der CDM-Ring möglicherweise nicht aus einem Einzeleinfall entstanden ist, sondern ebenso gut aus dem Einfall mehrerer Zwerggalaxien herrühren könnte. Sowohl das Modell des Einzeleinfalls als auch ein Modell des Einfalls dreier Zwerggalaxien kann die Rotationskurve der Milchstraße gut reproduzieren, siehe Abbildung \ref{fig:Rotationskurven}.

Ergebnisse wie die eben vorgestellten wurden bereits in anderen Arbeiten über andere Methoden und andere Modelle versucht zu erhalten.
Es gibt n-Teilchen Simulationen und semianalytische Modelle, welche die Entstehung eines CDM-Rings durch einen Einzeleinfall einer Zwerggalaxie modellieren, vergleiche zum Beispiel \citet{Kazantzidis08} und \citet{penarrubia05}. Auch diese Modelle gehen von Orbits kleiner Inklination und kleiner numerischer Exzentrizität aus. Allerdings wird in diesen Modellen nicht auf die mögliche Entstehung des Rings aus mehreren einfallenden Zwerggalaxien eingegangen. 

Das in dieser Arbeit vorgestellte semianalytische Modell ist also in der Lage, Aussagen über die Entstehung des CDM-Rings bei $R=14$ kpc in der Milchstraße zu machen, die sich mit den Ergebnissen von n-Teilchen-Simualtionen und anderer semianalytischer Modelle decken und diese, trotz der Einfachheit des hier vorgestellten Modells, ergänzen.

\subsection{Bezug zu anderen Galaxien}
Auch in Rotationskurven anderer Galaxien finden sich, wie bei der Milchstraße, Einbrüche der Rotationsgeschwindigkeit bei bestimmten Abständen zum galaktischen Zentrum. So zeigt NGC 660 (Hubble-Typ Sc) bei einem Abstand von etwa 10kpc zum galaktischen Zentrum einen Einbruch der Rotationsgeschwindigkeit, welcher verbunden mit einem Ring aus Sternen großer Inklination bezüglich der galaktischen Ebene ist \citep{honma97b}. Auch eine solche Struktur lässt sich mit dem hier vorgestellten Modell erklären, setzt jedoch einen Orbit der eingefallenen Zwerggalaxie hoher Inklination voraus. Auch NGC 1808 (Sbc) und NGC 4258 (Sb) zeigen niedrigere Rotationsgeschwindigkeit um einen Abstand von 10kpc zum galaktischen Zentrum, allerdings wird kein Sternring bei diesem Radius beobachtet. Auch hier können eingefallene Zwerggalaxien für die Einbrüche der Rotationsgeschwindigkeit verantwortlich sein, bei diesen Galaxien allerdings mit Orbits geringer Inklination, so dass die Sternströme innerhalb der galaktischen Ebene liegen. Das Finden ähnlicher Strukturen bei verschiedenen Galaxien deutet auf eine ähnliche Entstehungsgeschichte und bekräftigt das hier vorgestellte Modell nach dem Prinzip der hierarchischen Galaxienentstehung.

\section{Zusammenfassung}
Um die Bildung eines Rings aus kalter Dunkler Materie (CDM) innerhalb der Ebene der Milchstraße bei einem Abstand von $R=14$ kpc zum galaktischen Zentrum zu erklären, wurde ein semianalytisches Computermodell entwickelt. Dieses Modell beschreibt die Entstehung des CDM-Rings gemäß dem Prinzip der hierarchischen Galaxienentstehung durch den Einfall von Zwerggalaxien in die Milchstraße. 

Eine Zwerggalaxie befindet sich zu Beginn der Simulation bei einem Abstand $r_i$ zum galaktischen Zentrum der Milchstraße und besitzt ein Orbit mit der Anfangsgeschwindigkeit $v_i$, der Inklination $i_i$ und der numerischen Exzentrizität $e_i$. 
Durch dynamische Reibung verliert die Zwerggalaxie an Drehimpuls und Bewegungsenergie und bewegt sich im Gravitationspotential der Milchstraße in Richtung des galaktischen Zentrums. Erreicht die Zwerggalaxie den Abstand $r_t$ zum galaktischen Zentrum, übersteigen die durch das Gravitationspotential der Milchstraße wirksamen Gezeitenkräfte die Kräfte, welche die Zwerggalaxie zusammen halten. Die Zwerggalaxie wird daher am Radius $r_t$ zu einem Sternstrom (bestehend sowohl aus Sternen als auch aus CDM und Wasserstoffgas) zerrissen, der aufgrund der Ringform und der auf ein größeres Gebiet verteilten Materie unanfällig für dynamische Reibung des Halos ist. Daher behält der Sternstrom seinen Abstand zum galaktischen Zentrum in etwa bei.
Ist die Inklination des Orbits der eingefallenen Zwerggalaxie klein, so kann wie beschrieben die Bildung eines Rings aus CDM innerhalb der Ebene der Milchstraße erklärt werden. Bei größeren Inklinationen können sich Sternströme außerhalb der galaktischen Ebene bilden. Das vorgestellte Modell ermöglicht außerdem, Aussagen über die physikalischen Eigenschaften der Zwerggalaxie sowie ihres Orbits zu erhalten. Um in einer Zeit von maximal $13.7 \cdot 10^9$ Jahren einen Ring aus CDM innerhalb der galaktischen Scheibe und mit geringer Exzentrizität zu erhalten, kann, so zeigen die Simulationen, die anfängliche Exzentrizität $e_i$ des Orbits der eingefallenen Zwerggalaxie nicht größer als $\approx 0.2$ gewesen sein.
Bei einer Masse des CDM-Rings von $9 \cdot 10^{10}$ Sonnenmassen ist die Mindestmasse, um von kleinen $r_i < 100$ kpc bis zum Radius von $R=14$ kpc zu gelangen, deutlich geringer als die Ringmasse. Dies ermöglicht die Annahme, dass der CDM-Ring nicht aus einer großen sondern aus mehreren kleinen Zwerggalaxien entstanden ist. Die Computersimulationen zeigen, dass sowohl ein Einzeleinfall als auch der Einfall mehrerer Zwerggalaxien die beobachtete Rotationskurve der Milchstraße gut reproduzieren kann.

\section{Fazit}
\label{Sec:Fazit}
Das in dieser Arbeit vorgestellte semianalytische Computermodell ermöglicht Aussagen über die physikalischen Eigenschaften von Zwerggalaxien, die einen Ring aus kalter Dunkler Materie in der Milchstraße nach ihrer Zerstörung durch Gezeitenkräfte gebildet haben. Die Ergebnisse des vorgestellten Modells decken sich mit denen anderer Publikationen und ergänzen diese um die Möglichkeit der Ringbildung durch den Einfall mehrerer kleiner Zwerggalaxien.

\section*{Danksagung}
Besonderer Dank gilt Dr. Hans-Otto Carmesin als Leiter der AG für Astronomie \& Jugend-forscht am Athenaeum Stade für die vielen anregenden Diskussionen, seine stets hilfreichen Hinweise sowie seine konstruktiv kritischen Anmerkungen, welche für diese Arbeit richtungsweisend waren.
\pagebreak

\end{document}